\documentclass[aps,prl,showpacs,twocolumn]{revtex4}
	\usepackage[dvips]{graphics}
	\usepackage{amsfonts,amssymb,amsmath}
	\newcommand{\be}{\begin{equation}}
	\newcommand{\ee}{\end{equation}}
	\newcommand{\bea}{\begin{eqnarray}}
	\newcommand{\eea}{\end{eqnarray}} 

\begin{document} \title{Accelerating cosmologies from
	compactification} \preprint{hep-th/0303097}
	\preprint{DAMTP-2003-23} \author{Paul K. Townsend}
	\email{P.K.Townsend@damtp.cam.ac.uk} \author{Mattias
	N.R. Wohlfarth} \email{M.N.R.Wohlfarth@damtp.cam.ac.uk}
	\affiliation{Department of Applied Mathematics and Theoretical
	Physics, Centre for Mathematical Sciences, University of
	Cambridge, Cambridge CB3 0WA, U.K.}

	\begin{abstract}  A solution of the $(4+n)$-dimensional vacuum
	Einstein equations is found for which spacetime is
	compactified on an $n$-dimensional compact hyperbolic manifold
	($n\ge 2$) of time-varying volume to a flat four-dimensional
	FLRW cosmology undergoing a period of accelerated expansion in
	Einstein conformal frame. This shows that the `no-go' theorem
	forbidding acceleration in `standard' (time-independent)
	compactifications of string/M-theory does not apply  to
	`cosmological' (time-dependent) hyperbolic compactifications.
	\end{abstract} 
	\pacs{98.80.Cq, 11.25.Mj, 11.25.Yb, 98.80.Jk}

	\maketitle  

	Astronomical observations appear to show that the universe is
	not only expanding but is undergoing accelerated expansion,
	see e.g. \cite{Riess}. In addition, recent measurements of the
	cosmic microwave background provide support for the hypothesis
	of accelerated expansion in a much earlier inflationary
	cosmological epoch, see e.g. \cite{LB,Bennett}.  Although it
	is not difficult to find cosmological models that exhibit
	these features, one would wish any such model to be derivable
	from a fundamental, and mathematically consistent, theory that
	incorporates both gravity and the standard model of particle
	physics. Most current attempts to place the standard model
	within such a framework start from the ten or
	eleven-dimensional spacetime of superstring/M-theory, in which
	case one needs a compactification of ten or eleven dimensional
	supergravity in which an effective four-dimensional cosmology
	undergoes one or more periods of accelerated expansion.
	However, it has been shown that no such solution exists when
	the six or seven dimensional `internal' space is a
	time-independent non-singular compact manifold without
	boundary \cite{GWG,MN}. Three observations go into the
	derivation of this `no-go' theorem. The first is that
	accelerated expansion requires a violation of the
	strong-energy condition. This is the condition on the stress
	tensor that, given the Einstein equations, implies $R_{00}
	\geq 0$, but the acceleration of  a FLRW (homogeneous and
	isotropic) universe is positive if and only if $R_{00}$ is
	negative. The strong energy condition is violated in many
	four-dimensional supergravity theories but, and this is the
	second observation, it is not violated by either
	eleven-dimensional supergravity or any of the ten-dimensional
	supergravity theories that serve as effective field theories
	for a superstring theory. The third observation is generic to
	any compactification of the type specified in the theorem: if
	the higher-dimensional stress tensor satisfies the strong
	energy condition then so will the lower-dimensional stress
	tensor.

	Clearly, any attempt to derive a viable cosmology from
	string/M-theory must circumvent this no-go theorem, and this
	is possible in one of two ways. Either one rejects ten or
	eleven-dimensional supergravity as the relevant starting point
	or one relaxes one or more of the premises of the
	theorem. Attempts to circumvent the theorem by the addition of
	higher-derivative `quantum correction' terms to the
	supergravity action, or appeals to non-geometrical solutions
	of string theory with no classical analogue would fall into
	the former category, but we are not aware of any successful
	attempt along these lines. Here we shall assume that ten or
	eleven-dimensional supergravity {\sl is} the relevant starting
	point. The following options are now available. One can give
	up the compact condition on the internal space; this has the
	advantage that there are then known `compactifications' that
	allow accelerating four-dimensional cosmologies \cite{HW,pkt},
	but the disadvantage that the four-dimensional spectrum is
	continuous. As no good way around this problem has been found,
	or seems likely to be found \cite{GH}, we discard this
	possibility.  The possibility of an internal space that is
	compact but has a boundary can be considered a special case of
	an internal space with singular subspaces. This appears to be
	an attractive way to escape the no-go theorem because M-theory
	includes branes and boundaries on which the matter fields of
	the standard model are likely to be located.  However, any
	singular internal space that is the limit of a sequence of
	non-singular spaces would, by continuity, suffer from the same
	problems as non-singular internal spaces, so one would expect
	to have to consider non-resolvable singularities. We are not
	optimistic about this option, although it certainly deserves a
	full investigation.

	The only remaining option is to give up the condition of
	time-independence of the internal space. As we are concerned
	with cosmological solutions, which are intrinsically
	time-dependent, there is no good reason to suppose that the
	internal space is not also time dependent. From the
	four-dimensional perspective this amounts to the supposition
	that there are time-dependent scalar fields. In this case we
	need to confront an ambiguity in the choice of `conformal
	frame' for the metric: given a scalar field $\phi$ and a
	metric $g$ one can take \begin{equation} \tilde g_{\mu\nu} =
	e^{2\phi} g_{\mu\nu} \label{rescale} \end{equation}  as a new,
	conformally rescaled, metric.  The choice of conformal frame
	for which the four-dimensional gravitational action is of
	Einstein-Hilbert type, with no scalar-field-dependent
	multiplicative factor, defines the `Einstein frame' metric. If
	one insists on the Einstein conformal frame then it is not
	immediately obvious how time-dependence of the internal
	manifold helps. Unless the compactification generates a scalar
	potential, or a cosmological constant, the four-dimensional
	stress-tensor will still satisfy the strong energy
	condition. Thus, toroidal compactification (of the vacuum
	Einstein equations) can never yield an accelerating universe
	{\sl in Einstein frame}. This conclusion does not apply if the
	metric is not in Einstein conformal frame: let $g_{\mu\nu}$ be
	the four-metric of an FLRW cosmology in standard coordinates,
	and let the scalar field $\phi$ of (\ref{rescale}) depend only
	on the time coordinate $t$; then \begin{equation} \tilde
	R_{00}= R_{00} -3\left[\ddot \phi(t) + H(t)\dot\phi\right]
	\end{equation} where $H(t)$ is the Hubble `constant'. This
	shows that positivity of $R_{00}$ does not imply positivity of
	$\tilde R_{00}$.

	    This point is illustrated by the following Kasner-type
	    metric: \begin{equation} ds^2 = -dt^2 + t^{2\alpha}d{\bf
	    x}\cdot d{\bf x} + t^{2\beta} ds^2(T^n).  \end{equation}
	    This solves the (4+n)-dimensional vacuum Einstein
	    equations if \begin{equation} \alpha = {\frac{3 \pm
	    \sqrt{3n(n+2)}}{3(n+3)}}\, , \qquad \beta = {\frac{n \mp
	    \sqrt{3n(n+2)}}{n(n+3)}}.  \end{equation} The upper sign
	    yields a standard decelerating four-dimensional FLRW
	    spacetime. The lower sign yields an accelerating but {\sl
	    contracting} four-dimensional FLRW spacetime for $n \ge
	    2$. However, by taking $t \rightarrow (t_\infty -t)$ we
	    get \begin{equation} ds^2 = ds^2_4 + (t_\infty
	    -t)^{2\beta} ds^2\left(T^n\right) \end{equation} where
	    $ds^2_4$ is a flat FLRW spacetime with scale factor
	    \begin{equation} a(t) = (t_\infty -t)^\alpha.
	    \end{equation} As $\dot a>0$ and $\ddot a >0$ for
	    $t<t_\infty$ we have accelerated expansion. However
	    $ds^2_4$ is not the Einstein-frame metric. The
	    Einstein-frame four-metric yields a {\sl decelerating}
	    universe.

	    Non-Einstein conformal frames have the feature that the
	    effective Newton constant becomes time-dependent in
	    cosmological solutions. For this reason, among others,
	    what is really needed is a cosmological compactification
	    of ten or eleven-dimensional supergravity that yields a
	    four-dimensional FLRW universe undergoing accelerated
	    expansion in {\sl Einstein frame}. We have just argued
	    (assuming the absence of a scalar potential generated by
	    fields other than the metric) that this cannot be achieved
	    by any toroidal compactification and the same argument
	    applies to compactification on any Ricci-flat
	    space. However, we shall show, by example, that there
	    exist cosmological compactifications on Einstein spaces of
	    {\sl negative curvature} that yield accelerating
	    four-dimensional FLRW cosmologies in Einstein frame.

	Consider an $n$-dimensional compact Einstein manifold with
	metric \begin{equation} d\hat s^2_n = \hat g_{mn} dy^mdy^n
	. \end{equation}  We will take it to have constant {\sl
	negative} curvature $\kappa$, so that $n\ge2$ and
	\begin{equation} R(\hat g)_{mn} = -(n-1)\kappa^2 \hat
	g_{mn}\,.  \end{equation} Such spaces are obtained by
	identifying hyperbolic $n$-space under the action of a freely
	acting discrete subgroup of its $SO(1,n)$ isometry group. The
	identifications break all continuous isometries, so there will
	be no massless Kaluza-Klein vector fields resulting from
	compactification on this space.

		Now consider the following $(4+n)$-metric parametrized
	    by functions of time $S(t)$ and $K(t)$: \begin{equation}
	    ds^2 = e^{3nt/(n-1)} K^{-\frac{n}{n-1}}\, ds^2_E   +
	    e^{-6t/ (n-1)} K^{\frac{2}{n-1}} \, d\hat s^2_n
	    \label{metric} \end{equation} where \begin{equation}
	    ds^2_E = -S^6 dt^2 + S^2d{\bf x} \cdot d{\bf x}\, .
	    \end{equation} This metric solves the $(4+n)$-dimensional
	    vacuum Einstein equations if \begin{equation}  S(t) =
	    e^{-(n+2)t/2(n-1)} K^{\frac{n}{2(n-1)}}\, , \end{equation}
	    and \begin{equation} K(t) =
	    \frac{\sqrt{3(n+2)/n}}{(n-1)\kappa \sinh
	    (\sqrt{3(n+2)/n}\, |t|)}\, .  \end{equation} Note that
	    $\kappa$ has dimensions of inverse length, but an implicit
	    dimensionful constant appearing in these expressions has
	    been set to unity.

\begin{figure}[h!]  {\par\centering
	    \resizebox*{1.0\columnwidth}{!}
	    {\includegraphics{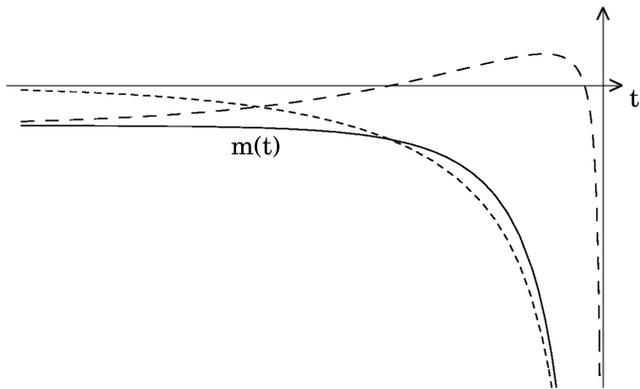}} \par}
	    \caption{\label{figure1}{\sl The function $m(t)$ for $n=7$
	    is compared to the sqare root of the right hand side of relation}
	    (\ref{acc}){\sl, in small dashes. The difference is
	    plotted in wide dashes and is positive in the interval of
	    acceleration.}}  \end{figure}

\begin{figure}[h!]  {\par\centering
	    \resizebox*{1.0\columnwidth}{!}{\includegraphics{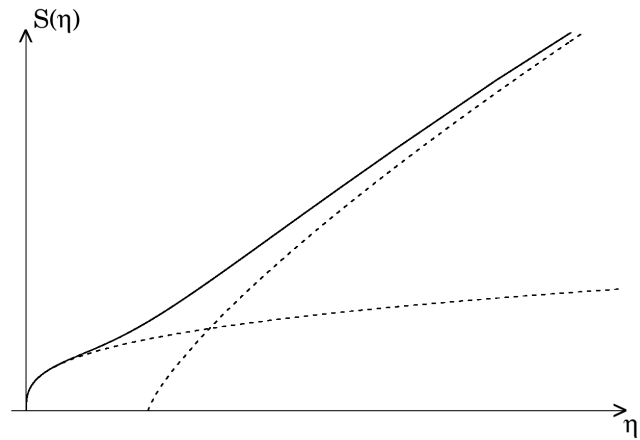}}
	    \par} \caption{\label{figure2}{\sl The scale factor
	    $S(\eta)$ of the four-dimensional universe is shown for
	    $n=7$ and $\kappa=1$. It clearly exhibits an accelerating
	    phase.}}  \end{figure}

	    The four-metric $ds^2_E$ is the {\sl Einstein frame}
	    metric of the four-dimensional theory that results from
	    the compactification of the $n$ internal dimensions. It
	    takes the standard FLRW form for a flat homogeneous
	    isotropic universe with scale factor $S$ in terms of the
	    time coordinate $\eta$ defined by \begin{equation} d\eta =
	    S^3(t) dt .  \end{equation} The four-dimensional universe
	    is expanding if $dS/d\eta >0$. This is equivalent to
	    $m(t)<0$ where \begin{equation} m(t) = 1+
	    \sqrt{\frac{3n}{n+2}} \coth \left(\sqrt{3(n+2)/n}\,\,
	    t\right) .  \end{equation} The universe undergoes
	    accelerated expansion if, in addition, $d^2S/d\eta^2
	    >0$. This is equivalent to \begin{equation}  \label{acc}
	    m(t)^2 < \frac{3(n-1)/(n+2)}{\sinh^2
	    \left(\sqrt{3(n+2)/n}\,\, t\right)}.  \end{equation}  Both
	    conditions are satisfied simultaneously for {\sl negative}
	    $t$ in a certain interval, as can be seen from the plot of
	    $m(t)$ in figure \ref{figure1} for the $n=7$ case of
	    relevance to M-theory compactifications.
	     It can be shown that the
	    universe is decelerating both as $\eta \rightarrow 0$
	    (corresponding to $t\rightarrow -\infty$) and as $\eta
	    \rightarrow \infty$ (corresponding to $t\rightarrow 0$
	    from $t<0$). Specifically, one finds that \begin{equation}
	    S \sim \eta^{1/3} \qquad \qquad (\eta \rightarrow 0)
	    \end{equation} which corresponds to the `stiff matter'
	    equation of state $p=\rho$ (for pressure $p$ and mass
	    density $\rho$), and \begin{equation} S \sim (\eta-
	    \eta_0)^{n/(n+2)} \qquad \qquad (\eta \rightarrow \infty)
	    \end{equation} which corresponds to the equation of state
	    $p= -[{\frac{n-4}{3n}}] \rho$; for $n=4$ this implies that
	    the universe ends as a dust-filled Einstein de-Sitter
	    universe, but for the $n=6,7$ cases of relevance to
	    M-theory compactifications the final epoch is one with
	    negative pressure matter, although the pressure is not
	    sufficiently negative for acceleration. These two
	    decelerating epochs are joined by an epoch of accelerated
	    expansion, as shown in figure \ref{figure2} for the $n=7$
	    case. The `matter' responsible for this behaviour is of
	    course the four-dimensional scalar field arising from the
	    Kaluza-Klein mode that scales the volume of the compact
	    internal space.  

	From these results one can see that the singularity of the
	function $K$ at $t=0$ is not a singularity of the
	Einstein-frame four-metric because $t=0$ is at an infinite
	proper time in the future of any event with $t<0$. It is also
	at an infinite proper time in the past of any event with
	$t>0$, so our solution really describes two possible
	universes, one for $t<0$ and another for $t>0$. Although the
	function $K$ is symmetric under $t\rightarrow -t$, the
	(4+n)-metric (\ref{metric}) is not, and only the $t<0$ case
	leads to a universe with an accelerating phase. Note that the
	$t<0$ and $t>0$ universes need not be (and are not) isometric;
	because $t=0$ is not in the spacetime, $t \rightarrow -t$ (in
	contrast to $\eta \rightarrow -\eta$) is not a transformation
	that reverses the cosmological flow.

	We have now shown how the no-go theorem of \cite{GWG,MN} may
	be circumvented by compactification on spaces with
	time-dependent metric. Such compactifications are, of course,
	typical of Kaluza-Klein cosmology, and have been extensively
	studied. However, no previous accelerating Kaluza-Klein
	cosmology that we are aware of can be considered as an escape
	from the no-go theorem of \cite{GWG,MN}, either because the
	strong energy condition is violated already in the higher
	dimension, or because the four-dimensional conformal frame is
	non-Einstein. Our solution has none of these undesirable
	features and yet not only has an accelerating phase,  but also
	has a built-in mechanism to both start and stop acceleration.

	Time-dependence of the internal metric was not in itself
	sufficient to yield an accelerating universe in four
	dimensions in Einstein frame. A hyperbolic compact internal
	space was also needed (because the analogous solution of the
	vacuum Einstein equations for an internal manifold of positive
	curvature does not allow acceleration). The absence of
	massless 
	Kaluza-Klein vector fields in hyperbolic compactifications
	would be a serious defect in a traditional Kaluza-Klein
	cosmology, but is an advantage from the modern perspective in
	which matter lives on space-filling branes. This also allows
	the matter to be supersymmetric and suggests a possible
	mechanism for supersymmetry breaking by coupling to the
	non-supersymmetric `bulk' gravitational theory arising from
	the supersymmetry breaking compactification (note that field
	theories with rigid supersymmetry can be coupled consistently,
	albeit non-supersymmetrically, to pure gravity).

	A discussion of the cosmological advantages of hyperbolic
	compactifications can be found in \cite{Kaloper,Trodden1}. One
	such advantage, emphasized in \cite{Kaloper} (where the
	relevant references to the mathematical literature may be
	found), arises from the remarkable fact that the only modulus
	of a compact hyperbolic Einstein space of dimension $n\ge3$ is
	its volume, so only the volume can be time-dependent; this
	means that there is no `rolling moduli' problem with this type
	of compactification. It seems remarkable that a model with so
	many attractive features can arise from a very simple
	compactification of M-theory.  \bigskip

{\bf Note added}: After having sent an earlier version of this paper
	to the archives we learned that the solution of the vacuum
	Einstein equations found here is the `zero-flux limit' of
	solutions of the non-vacuum Einstein equations obtained
	earlier in a different context by Chen {\it et al.}
	\cite{Chen} and Ohta \cite{Ohta}. This point has since been
	elaborated on in a number of papers
	\cite{Ohta2,Roy,Wohlfarth,GE,Chen2}. In the presence of flux,
	the relevant equations are not the vacuum Einstein equations
	and acceleration can also occur for compactification on spaces
	of non-negative curvature.

	 \bigskip

	    \begin{acknowledgments}  {\sl Acknowledgments.} We are
	    grateful to Gary Gibbons and Neil Turok for helpful
	    discussions. We also thank Ishwaree Neupane for e-mail
	    correspondence, which led us to make some minor
	    clarifications. MNRW acknowledges financial support from
	    the Gates Cambridge Trust.  \end{acknowledgments}


		\end{document}